\documentclass[prb,twocolumn,showpacs,floatfix,amsfonts]{revtex4}

\usepackage{graphicx,graphics,color}
\usepackage{bm}
\usepackage{amsmath}
\usepackage{amssymb}
\usepackage{epsfig}

\newcommand{\beq}{\begin{equation}}
\newcommand{\eeq}{\end{equation}}
\newcommand{\beqarray}{\begin{eqnarray}}
\newcommand{\eeqarray}{\end{eqnarray}}

\newcommand{\Hc}{\ensuremath{\mbox{H.c.}}} 
\newcommand{\Ham}[1][]{\ensuremath{{\cal{H}}_{\text{\tiny{#1}}}}} 
\newcommand{\eq}[1]{Eq.~(\ref{#1})} 
\newcommand{\fig}[1]{Fig.~\ref{#1}} 
\newcommand{\sgn}[2][]{\ensuremath{\text{sgn}_{#1}(#2)}} 

\begin{document}

\title{Competing orderings in an extended Falicov-Kimball model}
\author{P. M. R. Brydon,$^{1}$ Jian-Xin Zhu,$^{2}$ M. Gul\'{a}csi,$^{1}$ and A. R. Bishop$^{2}$}
\affiliation{$^{1}$ Department of Theoretical Physics, Institute
of Advanced Studies, Australian National University, Canberra, ACT
0200, Australia \\
$^{2}$ Theoretical Division, Los Alamos National Laboratory, Los
Alamos, NM 87545, U.S.A.}

\date{\today}

\begin{abstract}
We present a Hartree-Fock study of the Falicov-Kimball model
extended by both on-site and non-local hybridization. We examine the
interplay between excitonic effects and the charge-density wave
(CDW) instability known to exist at zero hybridization. It is found
that the CDW state remains stable in the presence of finite
hybridization; for on-site hybridization the Coulomb interaction
nevertheless strongly enhances the excitonic average above
its value in the non-interacting system. In contrast, for non-local
hybridization, we observe no such enhancement of the excitonic average
or a spontaneous on-site hybridization potential. Instead, we find
only a significant suppression of the excitonic correlations in
the CDW state. A phenomenological Ginzburg-Landau analysis is
also provided to understand the interplay.
\end{abstract}

\maketitle

The Falicov-Kimball model (FKM) describes a tight-binding system of
itinerant $d$-electrons interacting via on-site Coulomb repulsion
$U$ with localized $f$-electrons of energy $\epsilon_{f}$. The FKM was
originally introduced as a minimal model of valence transitions in
systems such as $\mbox{SmB}_6$ and Ce: by varying the inter-orbtial
Coulomb repulsion $U$ or the $f$-level $\epsilon_{f}$, both
discontinuous and continuous changes in the distribution of the
electrons across 
the localized and itinerant states were found.~\cite{FKMoriginal} It
was soon realised, however, that some overlap between the $d$- and
$f$-wavefunctions was an essential feature of most systems
displaying valence instabilities.~\cite{LRP81} This ``mixing'' of
the electron wavefunctions may be explicitly introduced by the
inclusion of a hybridization potential $V$. A variety of methods,
including Hartree-Fock,~\cite{L78} real-space renormalization
group~\cite{HH82} and alloy-analog approximation,~\cite{BC82}
revealed that the hybridization removed the previously observed
discontinuous valence transitions. Work on the FKM ceased in the
mid-1980s as it became apparent that the Periodic Anderson Model
offered a more realistic description of valence transition
physics.~\cite{C86}

As interest in the FKM as a model of valence transitions waned, it
was adopted as a model of a simple binary alloy.~\cite{KL/BS86} In
the limit of vanishing hybridization the $f$-electron occupation at
each site is a good quantum number: fixing the $f$- and
$d$-populations, the ground state is identified as the configuration
adopted by the $f$-electrons that minimizes the energy of the
conduction electrons. In particular, for a bipartite lattice at
half-filling and equal concentration of $d$- and $f$-electrons, the
$f$-electrons occupy the sites of one sublattice only, the so-called
chequerboard phase. For dimension $d\geq2$, this chequerboard
charge-density wave (CDW) state obtains for temperatures below a
critical temperature $T_{CDW}$; above this temperature a disordered
phase is realized. For $d=1$ the critical temperature is zero. We note
that the FKM as a binary alloy has been extensively studied in the
case of infinite dimension $d\rightarrow\infty$: the dynamical 
mean-field theory (DMFT) gives an exact solution in this limit.~\cite{FZ03} 

The FKM with hybridization has lately attracted renewed attention
due to the investigation of optical properties in this model by
Portengen \emph{et al.}~\cite{POS96} Following closely Leder's
Hartree-Fock (HF) work,~\cite{L78} they found that the Coulomb
repulsion induced an effective on-site hybridization; this effect
was sufficiently strong that it persists in the limit of negligible
hybridization. In fact, their calculations were performed
exclusively in this limit: their solution with non-zero polarization
or excitonic average $\langle{d^{\dagger}f}\rangle$ is
indistinguishable from the well-known excitonic insulator (EI)
state.~\cite{EI} The ``spontaneous'' excitonic average was
interpreted as evidence of electronic ferroelectricity. Their HF
solution, however, assumed a homogeneous ground state for the
system; the possibility of a CDW ground state was not considered.

The problem of reconciling the results of Portengen \emph{et al.}
with the known CDW instability has only been partially addressed.
Since the DMFT equations are no longer exactly solvable for
non-zero hybridization potential, Czycholl~\cite{C99} performed a
HF analysis for the $d\rightarrow\infty$ model. It was found that
for $V=0$ there was no spontaneous excitonic average and that the
CDW phase was stable against sufficiently small on-site
hybridization. For a given $U$ there  was a critical hybridization
$V_{c}(U)$ under which
 the CDW phase prevails. Czycholl nevertheless concluded
that the the inter-orbtial $U$ could strongly renormalize the
hybridization, and so could be important in the description of the
optical properties of strongly correlated electron systems. Also
working in the limit of large spatial dimensions, Zlati\'{c} 
{\it et al.} noted that for $V=0$ the hybridization susceptibility
diverges as $T\rightarrow0$, although they concluded that a
generalization of the FKM would be required for
$\langle{d^{\dagger}f}\rangle\neq{0}$ at finite
temperatures.~\cite{ZFLC01}

Comparatively little work has been done on this problem in finite
dimensions. Farka\v{s}ovsk\'{y} has used exact-diagonalization and
the density matrix renormalization group methods on small
one-dimensional systems to rule out the possibility of a spontaneous
excitonic average at zero temperature.~\cite{FNEF} By the same
methods, Farka\v{s}ovsk\'{y} has also analyzed the effect of
local~\cite{F97} and non-local~\cite{F04} hybridization; these works
are more concerned with the effect of the hybridization on valence
transitions, ignoring the possibility of an excitonic
renormalization of the hybridization potentials. Batista and
co-workers have claimed that a non-local hybridization stabilizes
ferroelectricity in a FKM extended by $f$-hopping;~\cite{Batista}
Sarasua and Continentino have investigated a similar
system.~\cite{SC04}

The FKM extended by hybridization cannot be solved exactly and
so it is necessary to use approximate methods to understand the
properties of the model. In this paper, we present a HF study of
the effect of the hybridization upon the CDW state on a
two-dimensional square lattice. The HF approximation is reliable for
small temperatures. It tends, however, to overestimate the stability
of ordered phases: in particular, the HF result for critical
temperature $T_{CDW}$ is very likely to be larger than the exact
value. Nevertheless, we can reasonably expect that the HF
approximation will give at least a qualitatively correct account of
the relative stability of ordered phases, even in 2D. The HF is
therefore an appropriate tool to study the competition between the EI
and CDW phases in the FKM. We consider only $\epsilon_{f}=0$ and
half-filling (the particle-hole symmetry point) as the CDW state here
adopts the simple chequerboard form; for these parameters also the
excitonic average takes its maximum as shown in the analysis of
Portengen {\it{et al.}}~\cite{POS96}

The FKM Hamiltonian for spinless Fermions is written
\beqarray \Ham &=&
-t\sum_{\langle{i,j}\rangle}d^{\dagger}_{i}d_{j} +
\epsilon_{f}\sum_{j}n^{f}_{j} +\sum_{i,j}\{V_{ij}d^{\dagger}_{i}f_{j}
+ \Hc\} \nonumber \\
&&+ U\sum_{j}n^{d}_{j}n^{f}_{j}\;. \label{eq:FKM}
\eeqarray
Some overlap between the $d$- and $f$-electron wavefunctions is
assumed, hence the hybridization term $V_{ij}$.
The concentration of electrons is fixed at $1=
\frac{1}{N}\sum_{j}\left\{\langle{n^{f}_{j}}\rangle+\langle{n^{d}_{j}}\rangle\right\}$
where $N$ is the number of sites. We measure all energies in terms of
the $d$-electron hopping integral $t$.

In our HF decoupling of the Coulomb interaction, we include the
possibility of the CDW state by allowing for a periodic modulation
of the order parameters:
\begin{eqnarray}
\langle{n^{f}_{j}}\rangle & = &
n^{f}+\delta_{f}\cos({\bf{Q}}\cdot{\bf{r}}_{j})\;, \label{eq:nf} \\
\langle{n^{d}_{j}}\rangle & = &
n^{d}+\delta_{d}\cos({\bf{Q}}\cdot{\bf{r}}_{j})\;, \label{eq:nc} \\
\langle{f^{\dagger}_{j}d_{j}}\rangle & = &
\Delta+\Delta_{Q}\cos({\bf{Q}}\cdot{\bf{r}}_{j})\;. \label{eq:Delta}
\end{eqnarray}
The nesting vector ${\bf{Q}}=(\frac{\pi}{a},\frac{\pi}{a})$ where
$a$ is the lattice constant. The order parameter of the CDW state is
$\delta_{d}$ and $\delta_{f}$ for the $d$- and $f$-electrons
respectively. Note that we require $\sgn{\delta_{f}} =
-\sgn{\delta_{d}}$. $\Delta$ is the excitonic average; in the
absence of an on-site hybridization potential $V$, $\Delta\neq0$
indicates the EI phase. When $V\neq{0}$, the EI-normal phase
transition is lifted from criticality, in analogy to the
ferromagnet-paramagnet transition in an external magnetic field. In
this case, we cannot speak of an EI phase, but rather an excitonic
enhancement of the hybridization. This will be apparent if $\Delta$
exceeds its value in the $U=0$ system. The modulation factor
$\Delta_{Q}$ is included in \eq{eq:Delta} for completeness. In the
usual HF treatment~\cite{L78,POS96} a homogeneous solution is
assumed and so $\delta_{d}=\delta_{f}=\Delta_{Q}=0$ for all values
of the Coulomb interaction.

We thus obtain for the HF Hamiltonian
\begin{eqnarray}
\Ham[HF] & = & -t\sum_{\langle{i,j}\rangle}d^{\dagger}_{i}d_{j} +
U\sum_{j}(n^{f}+\delta_{f}\cos({\bf{Q}}\cdot{\bf{r}}_{j}))n^{d}_{j}
\nonumber \\
&& +
U\sum_{j}(n^{d}+\delta_{d}\cos({\bf{Q}}\cdot{\bf{r}}_{j}))n^{f}_{j}
\nonumber \\
&&
+\sum_{ij}\biggl{\{}(V_{ij}-U\left[\Delta+\Delta_{Q}\cos({\bf{Q}}\cdot{\bf{r}}_{j})\right]
\delta_{ij})d^{\dagger}_{i}f_{j}\nonumber \\
&& +\Hc\biggr{\}}\;. \label{eq:HFHam}
\end{eqnarray}
An important feature of this Hamiltonian is the mean-field
renormalization of the $d$-$f$ hybridization potential by the
inter-orbital Coulomb interaction,
$V_{ij}\rightarrow{V_{ij}-U\left[\Delta+\Delta_{Q}\cos({\bf{Q}}\cdot{\bf{r}}_{j})\right]\delta_{ij}}$.
The effective on-site hybridization potential introduced by the
decoupling of the interaction is responsible for the spontaneous
polarization in Portengen {\it{et al.}}'s work.
$\Ham[HF]$ is diagonalized by the canonical transform
\beq
\gamma^{m}_{\bf{k}} = u^{m}_{\bf{k}}d_{\bf{k}} +
v^{m}_{\bf{k}}d_{\bf{k+Q}} + \xi^{m}_{\bf{k}}f_{\bf{k}} +
\zeta^{m}_{\bf{k}}f_{\bf{k+Q}}\;, \label{eq:CT}
\eeq
where $m=1,2,3,4$. The coefficients in \eq{eq:CT}
are obtained by solving the associated Bogoliubov-de Gennes (BdG)
eigenequations: \beq H_{\mathbf{k}} \Psi_{\mathbf{k}}^{m} =
E_{\mathbf{k}}^{m} \Psi_{\mathbf{k}}^{m}\;,
   \label{eq:BdG}
\eeq
where \beq H_{\mathbf{k}} = \left(\begin{array}{cccc}
\epsilon_{\bf{k}}+Un^{f} & U\delta_{f} & V_{\bf{k}}-U\Delta &
-U\Delta_{Q} \\
   U\delta_{f} & \epsilon_{\bf{k+Q}}+Un^{f} & -U\Delta_{Q} &
V_{\bf{k+Q}}-U\Delta \\
V_{\bf{k}}^{\ast}-U\Delta^{\ast} & -U\Delta_{Q}^{\ast} & Un^{d} &
U\delta_{d} \\
-U\Delta_{Q}^{\ast} & V_{\bf{k+Q}}^{\ast}-U\Delta^{\ast} & U\delta_{d}
& Un^{d}
\end{array}\right)
\eeq and \beq \Psi_{\mathbf{k}}^{m}= (u^{m}_{\bf{k}},
v^{m}_{\bf{k}}, \xi^{m}_{\bf{k}}, \zeta^{m}_{\bf{k}})^{Transpose}
\eeq Here $\epsilon_{\bf{k}}=-2t(\cos(k_{x}a)+\cos(k_{y}a))$ is the
$d$-electron energy dispersion. The self-consistency equations for
the HF parameters may be written in terms of the BdG eigenvectors:
\beqarray n^{d} &= & \frac{1}{N}\sum_{\bf{k}}{}^{\prime}
\left\{\langle{d^{\dagger}_{\bf{k}}d_{\bf{k}}}\rangle
+\langle{d^{\dagger}_{\bf{k+Q}}d_{\bf{k+Q}}}\rangle\right\}
\nonumber \\ &=&
\frac{1}{N}\sum_{\bf{k}}{}^{\prime}\sum_{m}\left\{u^{m\ast}_{\bf{k}}u^{m}_{\bf{k}}
+ v^{m\ast}_{\bf{k}}v^{m}_{\bf{k}}\right\}f(E^{m}_{\bf{k}})\;.
\label{eq:CTnd} \\
\delta_{d} &= & \frac{1}{N}\sum_{\bf{k}}{}^{\prime}
\left\{\langle{d^{\dagger}_{\bf{k+Q}}d_{\bf{k}}}\rangle
+\langle{d^{\dagger}_{\bf{k}}d_{\bf{k+Q}}}\rangle\right\} \nonumber
\\ &=&
\frac{1}{N}\sum_{\bf{k}}{}^{\prime}\sum_{m}\left\{v^{m\ast}_{\bf{k}}u^{m}_{\bf{k}}
+ u^{m\ast}_{\bf{k}}v^{m}_{\bf{k}}\right\}f(E^{m}_{\bf{k}})\;.
\label{eq:CTdeltad} \\
n^{f} &= & \frac{1}{N}\sum_{\bf{k}}{}^{\prime}
\left\{\langle{f^{\dagger}_{\bf{k}}f_{\bf{k}}}\rangle
+\langle{f^{\dagger}_{\bf{k+Q}}f_{\bf{k+Q}}}\rangle\right\}
\nonumber \\ &=&
\frac{1}{N}\sum_{\bf{k}}{}^{\prime}\sum_{m}\left\{\xi^{m\ast}_{\bf{k}}\xi^{m}_{\bf{k}}
+ \zeta^{m\ast}_{\bf{k}}\zeta^{m}_{\bf{k}}\right\}f(E^{m}_{\bf{k}})
\;.
\label{eq:CTnf} \\
\delta_{f} &= & \frac{1}{N}\sum_{\bf{k}}{}^{\prime}
\left\{\langle{f^{\dagger}_{\bf{k+Q}}f_{\bf{k}}}\rangle
+\langle{f^{\dagger}_{\bf{k}}f_{\bf{k+Q}}}\rangle\right\} \nonumber
\\ &=&
\frac{1}{N}\sum_{\bf{k}}{}^{\prime}\sum_{m}\left\{\zeta^{m\ast}_{\bf{k}}\xi^{m}_{\bf{k}}
+ \xi^{m\ast}_{\bf{k}}\zeta^{m}_{\bf{k}}\right\}f(E^{m}_{\bf{k}})\;.
\label{eq:CTdeltaf} \\
\Delta &= & \frac{1}{N}\sum_{\bf{k}}{}^{\prime}
\left\{\langle{f^{\dagger}_{\bf{k}}d_{\bf{k}}}\rangle
+\langle{f^{\dagger}_{\bf{k+Q}}d_{\bf{k+Q}}}\rangle\right\}
\nonumber \\ &=&
\frac{1}{N}\sum_{\bf{k}}{}^{\prime}\sum_{m}\left\{\xi^{m\ast}_{\bf{k}}u^{m}_{\bf{k}}
+ \zeta^{m\ast}_{\bf{k}}v^{m}_{\bf{k}}\right\}f(E^{m}_{\bf{k}})\;.
\label{eq:CTDelta} \\
\Delta_{Q} &= & \frac{1}{N}\sum_{\bf{k}}{}^{\prime}
\left\{\langle{f^{\dagger}_{\bf{k+Q}}d_{\bf{k}}}\rangle
+\langle{f^{\dagger}_{\bf{k}}d_{\bf{k+Q}}}\rangle\right\} \nonumber
\\ &=&
\frac{1}{N}\sum_{\bf{k}}{}^{\prime}\sum_{m}\left\{\zeta^{m\ast}_{\bf{k}}u^{m}_{\bf{k}}
+ \xi^{m\ast}_{\bf{k}}v^{m}_{\bf{k}}\right\}f(E^{m}_{\bf{k}})\;.
\label{eq:CTDeltaQ} \eeqarray The prime denotes summation over half
the Brillouin zone; $f(E)=1/\{1+\exp[\beta(E-\mu)]\}$ is the Fermi
distribution function. The chemical potential $\mu$ is determined by
the condition $1 =
N^{-1}\sum_{\bf{k}}{}^{\prime}\sum_{m}f(E^{m}_{\bf{k}})$. We use an
exact diagonalization method to solve the BdG
equation~(\ref{eq:BdG}) self-consistently. We start with an initial
set of order parameters. By solving Eq.~(\ref{eq:BdG}), the new
order parameters are computed via Eqs.~(\ref{eq:CTnd}) to
({\ref{eq:CTDeltaQ}) and are substituted back into
Eq.~(\ref{eq:BdG}). The iteration is repeated until a desired
accuracy is achieved.

We first consider the case of an on-site hybridization,
$V_{ij}=V\delta_{ij}$. In agreement with previous
work~\cite{C99,ZFLC01,FNEF} we find that for vanishing
hybridization the CDW phase is {\it{always}} stable against the EI
phase and there is no spontaneous excitonic average. The CDW order
displayed by the $V=0$ ground state will persist in the presence of
sufficiently small hybridization potentials, although the transition
temperature $T_{CDW}$ will be considerably suppressed,
see~\fig{fig:Onsite_CDW}. We find, however, that for finite
hybridization the Coulomb interaction will strongly enhance the
magnitude of $\Delta$. We plot the variation of $|\Delta|$ with
temperature in~\fig{fig:Onsite}. Note that since $\Delta=-|\Delta|$
there is a large renormalization of the hybridization potential due to
the mean-field decoupling of the Coulomb interaction~\eq{eq:HFHam}.
Comparing the homogeneous
solution without the CDW ordering (the solid line) with the solution
with a coexisting CDW ordering (the dotted line), we find a
significant suppression of $|\Delta|$ at the onset of the CDW order at
$T\sim{0.1t}$. Even within the CDW phase, however, the excitonic
enhancement of the on-site hybridization is still apparent as
$|\Delta|$ exceeds its value within the non-interacting system (the
dashed line in~\fig{fig:Onsite}). We do not find any evidence of
non-zero $\Delta_{Q}$.

\begin{figure}
\includegraphics[width=8cm]{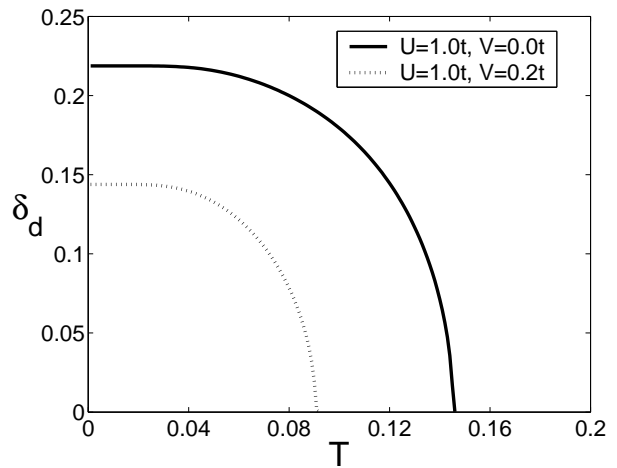}
\caption{\label{fig:Onsite_CDW}Variation of the CDW order parameter
   $\delta_{d}$ with temperature in the absence (solid line) and
   presence (dotted line) of an on-site hybridization potential.}
\end{figure}

\begin{figure}
\includegraphics[width=8cm]{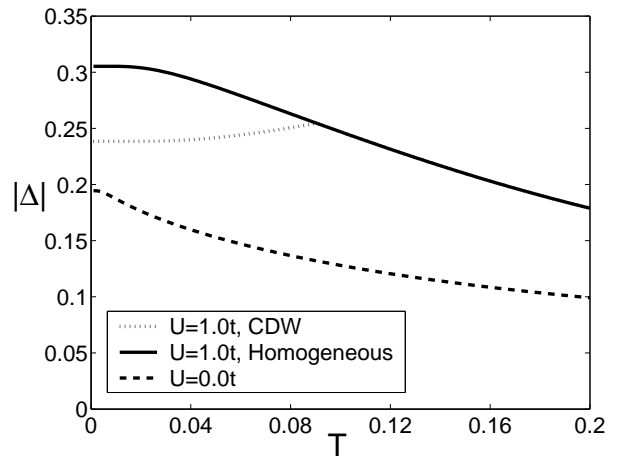}
\caption{\label{fig:Onsite}Comparison of the excitonic average
    $|\Delta|$ in the absence (solid line) and presence (dotted line) of
    the CDW instability for $U=1.0t$ with the value in the
    non-interacting system (dashed line) for $V=0.2t$.}
\end{figure}

This competition can be understood from a phenomenological
Ginzburg-Landau (GL) theory. The GL free energy density, in terms of
both the CDW ($\delta_d$) and EI ($\Delta$) order parameters, can be
constructed from a symmetry analysis:
\begin{eqnarray}
f&=&\alpha_{EI} \vert \Delta \vert^{2} + \alpha_{CDW } \vert
\delta_d \vert^{2} +\beta_{1}\vert \Delta \vert^{4}+\beta_{2}\vert
\delta_d \vert^{2}
\nonumber \\
&&+\beta_{3} \vert \Delta \vert^{2} \vert \delta_d \vert^{2}
-\beta_{4} (\Delta^{*2} \delta_{d}^{2} +\Delta^{2} \delta_{d}^{*2}
)\;,
\end{eqnarray}
where we assume $\alpha_{EI}=\alpha_{EI}^{\prime}(T-T_{EI}^{0})$ and
$\alpha_{CDW}=\alpha_{CDW}^{\prime}(T-T_{CDW}^{0})$. We assume
$\beta_i$ ($i=1,2,3,4$) are all positive. In the region where
$T_{EI}^{0}>T_{CDW}^{0}$,~\cite{Note}  the second phase transition
temperature for the CDW ordering is renormalized by the pre-existing
EI order parameter:
\begin{equation}
T_{CDW}=T_{CDW}^{0}-
\frac{(\beta_3-2\beta_4)(T_{EI}^{0}-T_{CDW}^{0})}
{2\beta_1\alpha_{CDW}^{\prime}/\alpha_{EI}^{\prime}-(\beta_3-2\beta_4)}\;.
\end{equation}
It means that when the EI order parameter pre-exists, the second phase
transition temperature for the appearance of the CDW order parameter
can be strongly suppressed by the dominant EI order parameter. This
explains why the transition temperature $T_{CDW}$ decreases with
increased hybridization potential $V$, as shown in
Fig.~\ref{fig:Onsite_CDW}. Below the second phase transition
temperature $T_{CDW}$, a little algebra yields
\begin{eqnarray}
\Delta &=& \biggl{[} \frac{-2\beta_2 \alpha_{EI} +
\alpha_{CDW}(\beta_3 -2\beta_4)}{4\beta_1\beta_2 -(\beta_3
-2\beta_4)^2}\biggr{]}^{1/2}\; \\
\delta_d &=& \biggl{[} \frac{-2\beta_1 \alpha_{CDW} +
\alpha_{EI}(\beta_3 -2\beta_4)}{4\beta_1\beta_2 -(\beta_3
-2\beta_4)^2}\biggr{]}^{1/2}\;.
\end{eqnarray}
Under the condition that the temperature derivative
$\alpha_{CDW}^{\prime}$ is larger than $\alpha_{EI}^{\prime}$, which
is indeed confirmed by our numerical results near $T_{CDW}$ (see
Figs.~\ref{fig:Onsite_CDW}-\ref{fig:Onsite}), $\alpha_{CDW}$ changes
more rapidly than $\alpha_{EI}$ when the temperature is lowered.
Consequently, the CDW order $\delta_d$ increases while the EI order
$\Delta$ decreases with the lowered temperature.

Despite the popularity of the on-site hybridization potential, this
is actually forbidden in real $d$-$f$ systems by parity
considerations.~\cite{C86} We are instead required  to consider a
non-local hybridization with inversion symmetry: the simplest such
potential is \beqarray V_{ij} &=&
t_{df}\biggl{[}\delta_{{i_x}j_x}(\delta_{{i_y}j_y+1} \nonumber \\
&&-\delta_{{i_y}j_y-1})+\delta_{{i_y}j_y}
(\delta_{{i_x}j_x+1}-\delta_{{i_x}j_x-1})\biggr{]}\;,\label{eq:IShyb}
\eeqarray where any site on the lattice is given by
${\bf{r}}_{i}=i_{x}a\hat{{\bf{x}}}+i_{y}a\hat{{\bf{y}}}$. This is a
particularly interesting case as the Coulomb-induced hybridization
has a different ($s$-wave) symmetry. In the non-interacting system,
the (on-site) excitonic average $\Delta$ vanishes; the non-local
hybridization potential instead gives rise to an anisotropic
excitonic average \beq \Xi =
\Im\left\{\frac{1}{N}\sum_{\bf{k}}\left(\sin(k_{x}a) +
\sin(k_{y}a)\right)\langle{f^{\dagger}_{\bf{k}}d_{\bf{k}}}\rangle\right\}\;.
\eeq The study of this quantity allows us to assess the effect of
the inter-orbital Coulomb repulsion upon the
$d$-$f$ hybridization.

\begin{figure}[th]
\includegraphics[width=8cm]{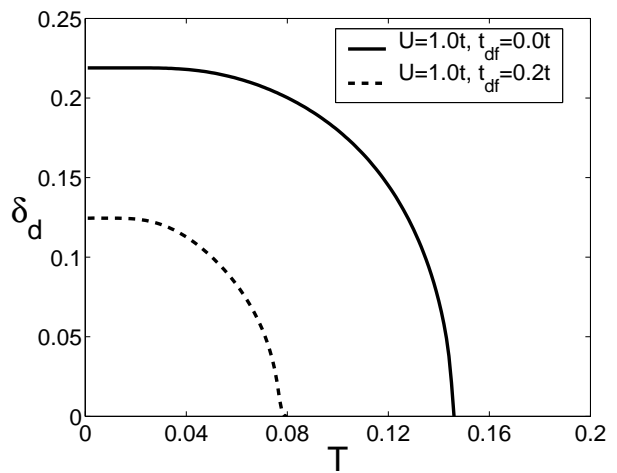}
\caption{\label{fig:IS_deltad}Variation of the CDW order parameter
    $\delta_{d}$ in the absence (solid line) and presence (dotted line)
    of a non-local hybridization potential.}
\end{figure}

\begin{figure}[th]
\includegraphics[width=8cm]{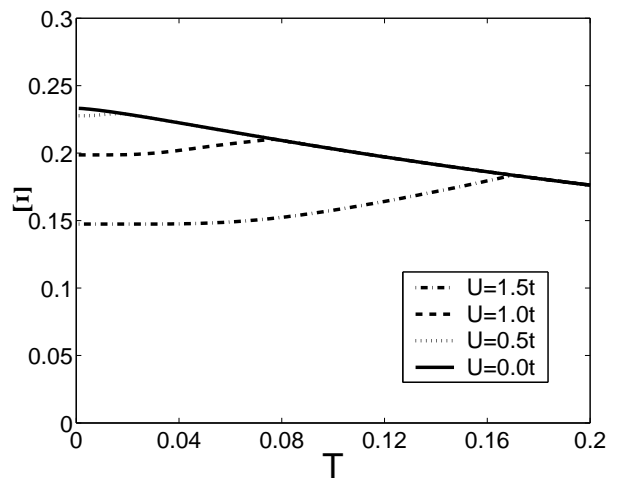}
\caption{\label{fig:IS}Comparison of the anisotropic exciton average
    $\Xi$ in the non-interacting and interacting systems. We have
    $t_{df}=0.2t$.}
\end{figure}

As with the on-site hybridization, we find that for given $U$ the
CDW phase is suppressed by the presence of the non-local
hybridization (see \fig{fig:IS_deltad}). We do not, however, find
any evidence for a Coulomb-induced on-site hybridization when the
CDW instability is allowed: for all non-zero $t_{df}$ we have
$\Delta=\Delta_{Q}=0$. Czycholl considered the appearance of an
on-site average $\Delta$ to be likely due to the substantial
excitonic enhancement of the on-site hybridization potential by the
Coulomb interaction.~\cite{C99} Our results clearly demonstrate that
this EI-like scenario, and the consequent formation of an electronic
ferroelectric state, is severely compromised by the presence of
non-local hybridization.

In~\fig{fig:IS} we plot $\Xi$ as function of temperature in both the
interacting and non-interacting systems. The onset of CDW order for
the given Coulomb values occurs at the point of intersection of the
broken lines with the non-interacting (solid) line. Remarkably, for
the standard homogeneous solution there is no effect on $\Xi$ due to
the Coulomb interaction: the variation of $\Xi$ with temperature
exactly follows the curve for the non-interacting system. Within the
CDW phase, however, $\Xi$ is suppressed below its value in the
non-interacting system. We offer the following explanation for this
anomaly: the hybridization potential \eq{eq:IShyb} connects the
A-sublattice $d$-orbitals with B-sublattice $f$-orbitals and vice
versa. Assume that in the CDW state the A-sublattice $d$-orbitals
have $n^{d}>0.5$ and so the B-sublattice $f$-orbitals have
$n^{f}>0.5$; clearly A-B sublattice $d$-$f$ hopping will be
suppressed, hence also the reduction in $\Xi$.

In conclusion, we have examined the competition between excitonic
and CDW instabilities in the FKM extended by both on-site
and non-local hybridization. In both cases, we
find that the CDW phase remains stable at low temperatures even in
the presence of a finite hybridization. For the local hybridization
we find that the Coulomb interaction nevertheless strongly
renormalizes the hybridization potential in agreement with previous
work.~\cite{C99} The situation is qualitatively different for the
more realistic non-local hybridization: there is no enhancement of
the non-local hybridization and the Coulomb interaction does not
induce a spontaneous on-site hybridization. Within the CDW phase,
the non-local hybridization is suppressed in line with the
increasing localization of the $d$- and $f$-electrons. The failure
of the Coulomb interaction to induce an effective on-site
hybridization except when such a term is already present casts
significant doubt over the
usefulness of~\eq{eq:FKM} as a minimal model for electronic
ferroelectricity. Inter-orbital Coulomb repulsion may nevertheless
still be important for understanding optical properties of
strongly-correlated electron systems: for example, a recent extension
of the FKM by $f$-electron hopping offers a plausible scenario where
the formation of an exciton BEC gives a spontaneous excitonic
average.~\cite{Batista} 

One of us (PMRB) acknowledges the hospitality of the Los Alamos
National Laboratory where part of this work was carried out. This
work was supported by the US DOE (JXZ and ARB).


\begin{thebibliography}{99}

\bibitem{FKMoriginal}L. M. Falicov and J. C. Kimball,
Phys. Rev. Lett. {\bf{22}}, 997 (1969); R. Ramirez and L. M. Falicov,
Phys. Rev. B {\bf{3}}, 2425 (1971).

\bibitem{LRP81}J. M. Lawrence, P. S. Riseborough and R. D. Parks,
Rep. Prog. Phys. {\bf{44}}, 1 (1981).

\bibitem{L78}H. J. Leder, Solid State Commun. {\bf{27}}, 579 (1978).

\bibitem{HH82}W. Hanke and J. E. Hirsch, Phys. Rev. B {\bf{25}}, 6748
    (1982).

\bibitem{BC82}E. Baeck and G. Czycholl, Solid State Commun. {\bf{43}},
    89 (1982).

\bibitem{C86}G. Czycholl, Phys. Rep. {\bf{143}}, 277 (1986).

\bibitem{KL/BS86}T. Kennedy and E. H. Lieb, Physica A {\bf{138}}, 320
    (1986); U. Brandt and R. Schmidt, Z. Phys. B {\bf{63}}, 45 (1986).

\bibitem{FZ03}J. K. Freericks and V. Zlati\'{c},
  Rev. Mod. Phys. {\bf{75}}, 1333 (2003), and references therein.

\bibitem{POS96}T. Portengen, Th. \"{O}streich and L. J. Sham,
Phys. Rev. B {\bf{54}}, 17452 (1996).

\bibitem{EI}L. V. Keldysh and Y. V. Kopaev, Sov. Phys. Solid State
    {\bf{6}}, 2219 (1965).

\bibitem{C99}G. Czycholl, Phys. Rev. B {\bf{59}}, 2642 (1999).

\bibitem{ZFLC01}V. Zlati\'{c} {\it{et al.}}, Philos. Mag. B {\bf{81}},
  1443 (2001).

\bibitem{FNEF}P. Farka\v{s}ovsk\'{y}, Phys. Rev. B {\bf{59}}, 9707 (1999);
 Phys. Rev. B {\bf{65}}, 081102(R) (2002).

\bibitem{F97}P. Farka\v{s}ovsk\'{y}, Z. Phys. B {\bf{104}}, 553 (1997).

\bibitem{F04}P. Farka\v{s}ovsk\'{y}, Phys. Rev. B {\bf{70}}, 035117 (2004).

\bibitem{Batista}C. D. Batista, Phys. Rev. Lett. {\bf{89}}, 166403
  (2002); C. D. Batista {\it{et al.}}, Phys. Rev. Lett. {\bf{92}},
  187601 (2004).

\bibitem{SC04}L. G. Sarasua and M. A. Continentino, Phys. Rev. B
  {\bf{69}}, 073103 (2004).

\bibitem{Note} Note that for a finite hybridization
potential the expectation value of the particle-hole pair operator,
$\langle f^{\dagger} d\rangle$ is a quantity characterizing the
charge polarization, not the spontaneous EI order. As such, the
characteristic temperature $T^{0}_{EI}$ is not a real EI phase
transition temperature.

\end{thebibliography}
\end{document}